\documentclass[10pt,aps,prd,nofootinbib,amsfonts,amssymb,amsmath,preprintnumbers,notitlepage,superscriptaddress,longbibliography,floatfix]{revtex4-1}
\pdfoutput=1
\usepackage{amsthm}
\usepackage{aas_macros}
\usepackage[utf8]{inputenc}
\usepackage[british]{babel}
\usepackage[Symbolsmallscale]{upgreek}
\usepackage[protrusion=true,tracking=true,kerning=true,spacing=true,final,babel=true]{microtype}
\usepackage{physics}
\usepackage{xcolor}
\usepackage{graphicx}
\usepackage[final]{hyperref}
\usepackage{nth}
\usepackage{cancel}
\usepackage{enumerate}
\usepackage[normalem]{ulem}

\def\be{\begin{equation}}
\def\ee{\end{equation}}
\def\bi{\begin{itemize}}
\def\ei{\end{itemize}}
\def\ben{\begin{enumerate}}
\def\een{\end{enumerate}}

\begin{document}
\title{Common-spectrum process versus cross-correlation for gravitational-wave
searches using pulsar timing arrays}

\author{Joseph~D.~Romano}
\affiliation{Department of Physics and Astronomy, Texas Tech University, 
Lubbock, TX 79409-1051, USA}
\email{joseph.d.romano@ttu.edu}

\author{Jeffrey~S.~Hazboun}
\affiliation{University of Washington Bothell, 18115 Campus Way NE, Bothell, WA 98011, USA}

\author{Xavier Siemens}
\affiliation{Department of Physics, Oregon State University, Corvallis, OR 97331, USA}

\author{Anne M.~Archibald}
\affiliation{School of Mathematics, Statistics, and Physics, Newcastle University, NE17RU UK}

\date{\today}

\begin{abstract}
The North American Nanohertz Observatory for Gravitational Waves
(NANOGrav) has recently reported strong statistical evidence
for a common-spectrum red-noise process for all pulsars, 
as seen in their 12.5-yr
analysis for an isotropic stochastic gravitational-wave 
background.
However, there is currently very little evidence for quadrupolar 
spatial correlations across the pulsars in the array, which is
needed to make a confident claim of detection of a stochastic
background.
Here we give a frequentist analysis of a very simple signal+noise 
model showing that the current lack of evidence for spatial 
correlations is consistent with the magnitude of the 
correlation coefficients for pairs of Earth-pulsar baselines
in the array, and the fact that pulsar timing arrays
are most-likely operating in the intermediate-signal regime.
We derive analytic expressions that allow one to compare the 
expected values of the signal-to-noise ratios for both the 
common-spectrum and cross-correlation estimators.
\end{abstract}

\maketitle

\section{Introduction}
\label{s:intro}

A stochastic gravitational-wave background (GWB) is 
formed from the superposition of GW signals 
that are either too weak or too numerous to individually 
detect~\cite{m00, christensen2018}.
Although we have not yet detected a stochastic background, we
know from the advanced LIGO-Virgo detections of individual
stellar-mass compact binaries~\cite{gw150914, gw170817, gwtc2, gwtc2}
that such a signal exists.
So it is just a matter of time before we reach the sensitivity 
level needed to make a confident detection of the corresponding
unresolved GWB signal.
For the astrophysical GWB produced by stellar-mass binary 
black holes (BBHs) throughout the universe, 
we expect to reach this level by the time the advanced LIGO~\cite{aLIGO2014}
and Virgo~\cite{aVirgo2014} detectors are operating at design sensitivity 
(in a couple of years time)~\cite{gw150914stoch, gw170817stoch}.
For the analogous GWB produced by supermassive 
($\sim\!10^9~M_\odot$) binary black holes (SMBBHs) associated
with galaxy mergers throughout the universe~\cite{jb03}, 
we may actually be a bit closer
to reaching that sensitivity level~\cite{rsg2015, tve+16}.

The North American Nanohertz Observatory for Gravitational Waves
(NANOGrav)~\cite{ransom+19} 
has recently reported strong statistical evidence for a 
common-spectrum red-noise process for all pulsars, 
as seen in their 12.5-yr
analysis for an isotropic stochastic signal~\cite{abb+2020}.
But there is currently very little evidence for quadrupolar 
spatial correlations across the pulsars~\cite{hd83}, which
is needed to make a confident claim of detection of
a stochastic gravitational-wave background.
As noted in their paper, the current lack of evidence for
spatial correlations is consistent with the reduction in 
signal power that comes from the magnitude of the 
correlation coefficients for pairs of Earth-pulsar baselines
being less than or equal to 0.5.
Here we give an explicit proof of that statement in the 
context of a very simple signal+noise model.
Using simple frequentist estimators of the amplitude of the GWB, 
we show that the common-spectrum (i.e., auto-correlation) 
estimator has a larger signal-to-noise ratio than a 
cross-correlation estimator in the 
{\em intermediate-signal regime}~\cite{sejr13}, 
even if there is no cross-correlated noise to compete against.
The estimators that we use have similarities to various frequentist GWB 
statistics in the pulsar timing literature \cite{abc+2009, ccs+2015, rsg2015}, 
but are purposefully simplified in order to highlight the 
difference between the auto-correlated and cross-correlated signal.

This simplification of the model leads to a few subtleties in how this model is
related to the actual situation.  
In the simple toy model that we shall describe below, 
the detection of a red-noise process of any kind would establish the presence of a
GWB (in technical terms, the measured signal-to-noise ratio relative to the 
white-noise-only null hypothesis would be sufficiently greater than zero).
But in reality, pulsars have intrinsic red
noise that must be distinguished from the common-mode effects of the GWB. 
While some red noise has long since been clearly
detected in many pulsars, an analysis of the properties of the red noise
is necessary to detect a common-mode process. 
We therefore define our signal-to-noise ratios relative to their variance 
{\em in the presence of a potentially non-zero signal}, 
thus taking into account the variance 
associated with the GWB itself~\cite{rsg2015}.
This is in contrast to the optimal 
statistic~\cite{ccs+2015}, which acts as a detection statistic by defining the noise
using the characteristics of a noise-only model.
While both of these statistics generically include intrinsic pulsar red noise, we omit this
complication for clarity, essentially assuming we have modeled the intrinsic red noise perfectly.
The argument of this paper is then
that with the same data, in the intermediate-signal regime where modern 
pulsar timing arrays operate~\cite{sejr13}, 
one can estimate the amplitude of a GWB using the auto-correlations 
more precisely than one can estimate it using 
the cross-correlations. This is essentially what is happening with the NANOGrav
12.5-yr data set, where the auto-correlations can be measured well enough to
detect a common-spectrum red-noise process, but the cross-correlations alone cannot 
yet be measured well enough to provide the unique indication that this is a
gravitational-wave signal.

The rest of the paper is organized as follows:
In Sec.~\ref{s:toy_model}, we define our simple signal+noise 
model, and in Sec.~\ref{s:optimal_estimators} we 
calculate the expected signal-to-noise ratios of the amplitude
of a potential GWB for the optimal common-spectrum 
and cross-correlation estimators, in both the weak-signal
and intermediate-signal regimes.
Our main finding is that the optimal common-spectrum estimator 
has a larger expected signal-to-noise ratio than the
optimal cross-correlation estimator in the intermediate-signal regime, 
in broad agreement with the results of NANOGrav's 12.5-yr analysis.
We conclude in Sec.~\ref{s:discussion} with a brief discussion,
comparing the main results obtained here 
to expectations for ground-based detectors, and more 
sophisticated statistical analyses, like
those actually performed on the 12.5-yr data~\cite{abb+2020}.

\section{Toy model}
\label{s:toy_model}

We consider a pulsar timing array consisting of
$N$ pulsars,
with uncorrelated white Gaussian noise described by power
spectra
\be
P_{n_a}(f) = 2\sigma_a^2\,\Delta t\,,
\qquad a=1,2,\cdots, N\,,
\ee
and a GWB with total auto-correlated power
\be
P_{\rm gw}(f) 
=\frac{A_{\rm gw}^2}{12\pi^2}\left(\frac{f}{f_{\rm yr}}
\right)^{2\alpha}f^{-3},
\qquad {\alpha=-2/3}\,,
\label{e:Pgw}
\ee
where $\alpha=-2/3$ is the spectral index appropriate 
for GW-driven emission from inspiraling SMBBHs~\cite{p01}.
In the above expressions, $P(f)$ are one-sided power
spectral densities for the timing residuals 
(units of $P(f)$ are ${\rm s}^2/{\rm Hz}$);
$\sigma_a^2$ are the noise variances 
for the pulsars (not necessarily equal to one another); 
$\Delta t$ is the sampling period
(typically one to three weeks for pulsar timing observations);
and $f_{\rm yr}\equiv 1/{\rm yr}$ is a reference 
frequency for evaluating the GWB amplitude.

\begin{figure}[h!tbp]
\centering
\includegraphics[width=0.5\textwidth]{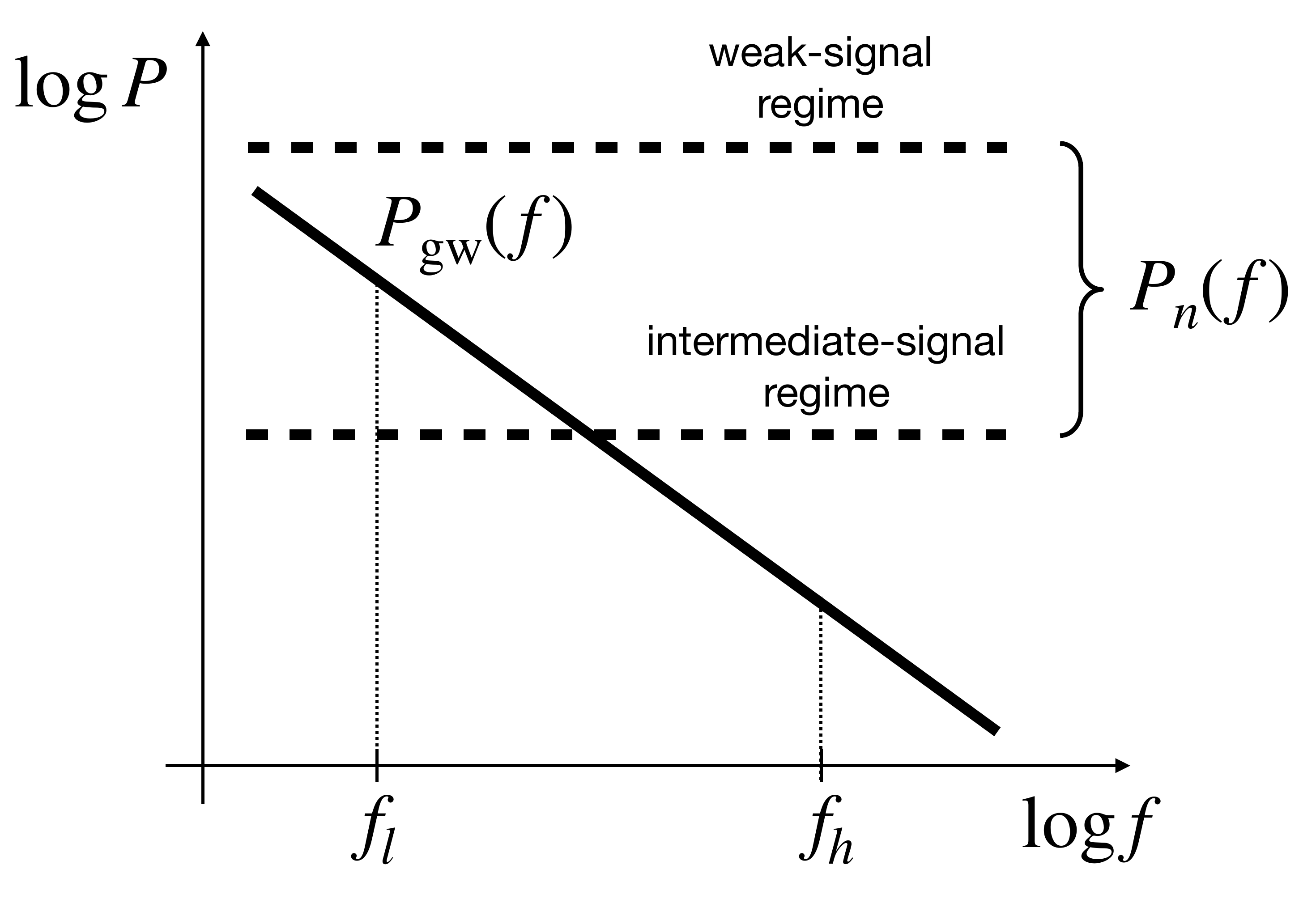}
\caption{Power spectra for both the GWB and
pulsar white noise, allowing for different levels of 
pulsar noise.
For all cases, the pulsar white noise
dominates the power in the GWB at high frequencies,
$P_n(f_h)\gg P_{\rm gw}(f_h)$.
In the weak-signal regime 
$P_{\rm gw}(f_l)\ll P_{n}(f_l)$,
while the intermediate-signal regime has
$P_{\rm gw}(f_l)\gg P_{n}(f_l)$.
For our toy model, we consider only two discrete
frequencies $f_l$ and $f_h$.}
\label{f:signalNoise}
\end{figure}

For simplicity, we will assume that there are only 
two frequency bins
\be
f_l\equiv f_{\rm low}\,,\qquad
f_h\equiv f_{\rm high}\,,
\ee
and that the pulsar white noise for all pulsars 
is much larger than the power in the GWB at high frequencies,
so that $P_{n_a}(f_h)\gg P_{\rm gw}(f_h)$.
At low frequencies, the pulsar white noise can be either
much greater or much less than the power in the GWB.
We call these two regimes the {\em weak-signal} 
and {\em intermediate-signal} regimes, respectively~\cite{sejr13}
(Figure~\ref{f:signalNoise}):
\begin{align}
&P_{\rm gw}(f_l)\ll P_{n_a}(f_l)
\quad\&\quad 
P_{\rm gw}(f_h)\ll P_{n_a}(f_h)
&({\rm weak\ signal\ regime})\,,
\\
&P_{\rm gw}(f_l)\gg P_{n_a}(f_l)
\quad\&\quad 
P_{\rm gw}(f_h)\ll P_{n_a}(f_h)
&({\rm intermediate\ signal\ regime})\,.
\label{e:regimes}
\end{align}
The unknown signal and noise parameters are 
$\sigma_a^2$ and $A_{\rm gw}^2$.
But to simplify the analysis, we will work with 
{\em rescaled} parameters 
$\bar\sigma_a^2$ and
$\bar A_{\rm gw}^2$ so that
\be
P_{n_a}(f) = \bar\sigma_a^2\,,
\quad
P_{\rm gw}(f) = \bar{A}{}_{\rm gw}^2\, f^{-13/3}\,,
\quad a=1,2,\cdots, N\,.
\ee

\begin{figure}[h!tbp]
\centering
\includegraphics[width=0.5\textwidth]{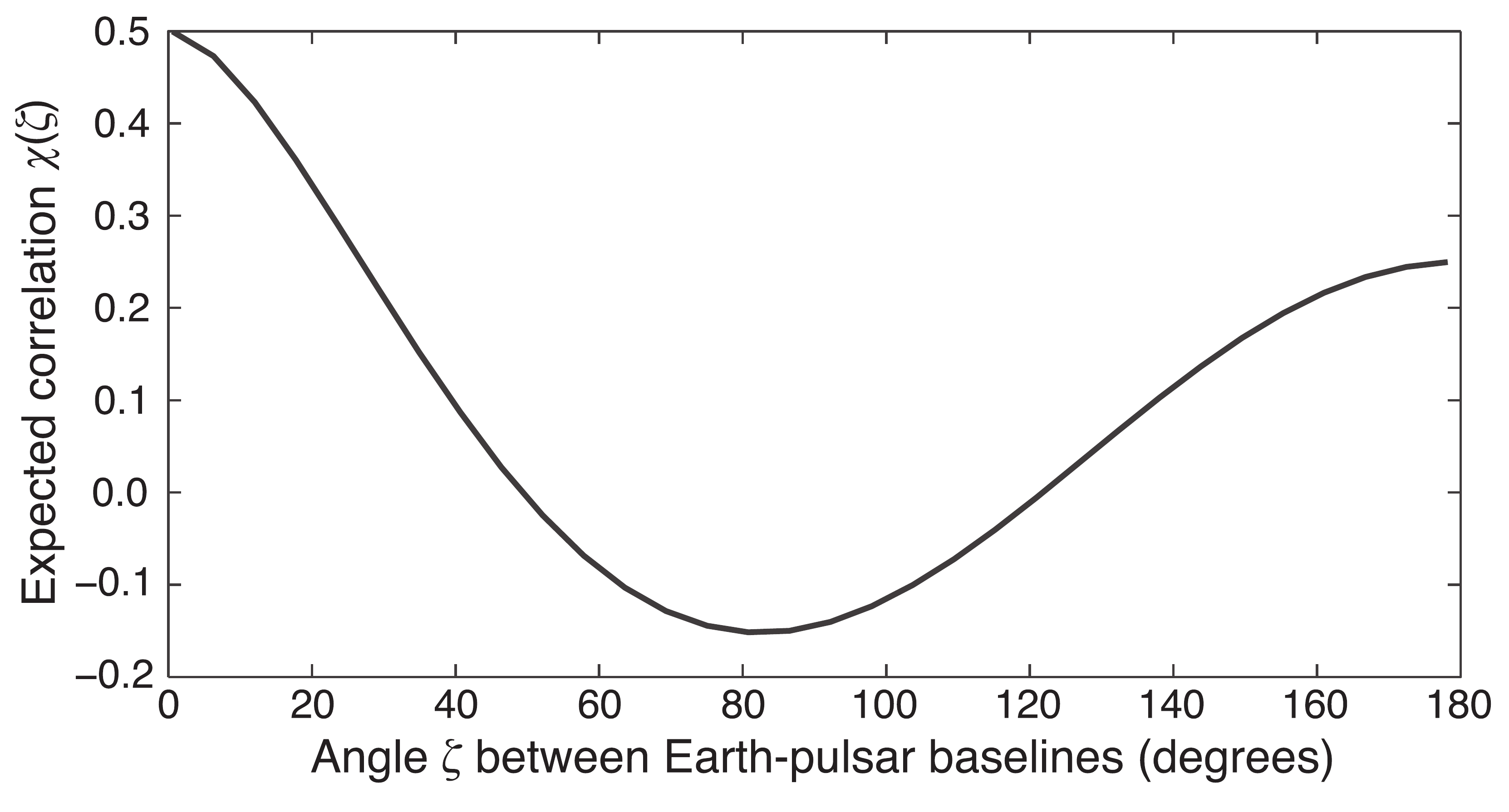}
\caption{Hellings-and-Downs curve, which gives the 
expected correlation between the timing residuals 
measured by two Earth-pulsar baselines as a function
of the angle $\zeta$ between the two baselines.}
\label{f:HD_curve}
\end{figure}

The timing-residual data for the $N$ pulsars can be 
written in the frequency domain as
\be
\tilde d_a(f) = \tilde h_a(f) + \tilde n_a(f)\,,
\qquad a=1,2,\cdots,N\,,
\ee
where the tildes denote Fourier transforms of the 
corresponding time-series.
The data covariance matrix is given by
\be
\tilde C(f)
=\left(
\begin{array}{cccc}
P_1(f) & P_{12}(f) & \cdots & P_{1N}(f)\\
P_{21}(f) & P_2(f) & \cdots & P_{2N}(f)\\
\vdots & \vdots & \ddots & \vdots \\
P_{N1}(f) & P_{N2}(f) & \cdots & P_N(f) \\
\end{array}
\right)\,,
\label{e:cov_power}
\ee
where
\be
P_a(f) \equiv P_{n_a}(f) + P_{\rm gw}(f)\,,
\quad
P_{ab}(f) = P_{ba}(f) \equiv \chi_{ab}\, P_{\rm gw}(f)\,,
\label{e:power_spectra}
\ee
and 
$\chi_{ab}\equiv \chi(\zeta_{ab})$ is the value of the
Hellings-and-Downs correlation function~\cite{hd83}
evaluated at the angular separation $\zeta_{ab}$ between two 
Earth-pulsar baselines
(Figure~\ref{f:HD_curve}).
Note that we are assuming here that there is no 
cross-correlated noise, so that only the GWB power 
spectrum appears in the off-diagonal terms, multiplied
by the Hellings-and-Downs coefficients.

It is also important to emphasize that $P_{\rm gw}(f)$
is the {\em total} 
auto-correlated power induced by the GWB in the 
timing residuals for each Earth-pulsar baseline.
This auto-correlated power has contributions 
from both the {\em Earth-term} and {\em pulsar-term} 
components of the GWB timing-residual response~\cite{hd83}, 
while only the 
Earth-term components of the response are correlated 
across pulsars in the array and contribute to $\chi(\zeta)$.
This is the reason for the factor of $0.5$ in the
normalization of the Hellings-and-Downs curve (Figure~\ref{f:HD_curve})
at zero angular separation.
The total auto-correlated power corresponds to a {\em single} 
Earth-pulsar baseline,
while the Hellings-and-Downs curve gives the expected correlation
coefficient for {\em two distinct} Earth-pulsar baselines,
even though the lines-of-sight to the two pulsars might be the same.

The standard estimators for the total 
auto-correlated and cross-correlated power spectra are
\be
\hat P_a(f) \equiv \frac{2}{T}|\tilde d_a(f)|^2\,,
\qquad
\hat P_{ab}(f) \equiv \frac{2}{T}{\rm Re}[\tilde d_a(f)\tilde d_b^*(f)]\,,
\ee
where $T$ is the total observation time, and the factor 
of 2 has been included to make these one-sided power 
spectrum estimates.
Since we only have two frequencies to work with, we
have $2N$ complex data points
\be
\tilde d_{al}\equiv 
\tilde d_a(f_l)\,,
\quad
\tilde d_{ah}\equiv 
\tilde d_a(f_h)\,,
\qquad a=1,2,\cdots, N\,,
\ee
from which to construct estimators for the $N+1$ parameters
$\bar\sigma_a^2$ and $\bar A_{\rm gw}^2$.

Given the inequalities in \eqref{e:regimes}, it immediately
follows that the white noise variances $\bar\sigma_a^2$,
can be estimated simply from 
the high-frequency estimators of the auto-correlated power:
\be
\hat{\bar\sigma}_a^2 \equiv \hat P_a(f_h)
=\frac{2}{T}|\tilde d_{ah}|^2\,, 
\qquad a=1,2,\cdots, N\,.
\ee
Similarly,
the amplitude $\bar A_{\rm gw}^2$ of the power in 
the GWB can be estimated from either the difference between the 
auto-correlated power at low frequency and the white noise
estimators at high frequency:
\be
\hat 
A_a^2\,f_l^{-13/3}
\equiv \hat P_a(f_l) -\hat{\bar\sigma}^2_a=
\frac{2}{T}\left(|\tilde d_{al}|^2 - |\tilde d_{ah})|^2\right)\,,
\qquad a=1,2,\cdots, N\,, 
\label{e:CP}
\ee
or the cross-correlated power at low frequencies
\be
\hat A_{ab}^2\,f_l^{-13/3}
\equiv {\hat P_{ab}(f_l)}/{\chi_{ab}}
=
\frac{2}{T}{{\rm Re}[\tilde d_{al}\tilde d_{bl}^*]}/{\chi_{ab}}\,,
\qquad ab=12, 13, \cdots \,,
\label{e:cross}
\ee
where the number of distinct pulsar pairs $ab$
is given by $N_{\rm pairs} \equiv N(N-1)/2$.
(We do not consider cross-correlation estimators evaluated at 
high frequencies, since the power in the GWB at high frequencies
is less than that at low frequencies by a large factor,
$(f_h/f_l)^{13/3}$.)
We will call the estimators $\hat A_a^2$, which are
constructed from the auto-correlated power spectrum estimates,
{\em common-spectrum} or common-process estimators.
Similarly, we will call the estimators $\hat A_{ab}^2$, which are
constructed from the cross-correlated power spectrum estimates,
{\em cross-correlation} or Hellings-and-Downs estimators.
Note that in all cases, the expectation value of these
estimators (averaged over different signal and noise realizations)
is just the amplitude of the power in the GWB:
\be
\langle \hat A_a^2\rangle 
=\langle \hat A_{ab}^2\rangle 
= \bar A_{\rm gw}^2\,.
\ee
We can further construct {\em optimal} 
(i.e., minimal-variance, unbiased) estimators of
$\bar A_{\rm gw}^2$ by forming appropriate linear 
combinations of these individual estimators.
Since we are interested in this paper in comparing
the common-spectrum and cross-correlation estimators,
we will define and analyze these two estimators separately in the
following section.

\section{Optimal estimators}
\label{s:optimal_estimators}

To construct optimal linear combinations of the
individual common-spectrum and cross-correlation
estimators defined in the previous section, 
we need to calculate
the covariance matrices of these estimators, 
from which we obtain the optimal weighting factors.
This is a consequence of a general result, which
says that given $N$ estimators $\hat X_i$ of
some quantity $X$, the optimal linear combination 
of the $\hat X_i$ is given by the weighted average
\be
\hat X_{\rm opt} 
\equiv \frac{\sum_i \lambda_i \hat X_i}
{\sum_j \lambda_j}\,,
\qquad
\lambda_i\equiv \sum_j \left(C^{-1}\right)_{ij}\,,
\ee
where $C_{ij}$ are the components of the covariance 
matrix for the estimators:
\be
C_{ij} \equiv \langle \hat X_i \hat X_j\rangle 
- \langle \hat X_i\rangle
\langle \hat X_j\rangle\,.
\ee
It is easy to show that the expected value of 
$\hat X_{\rm opt}$ is $X$ 
(since $\langle\hat X_i\rangle=X$ and 
$X_{\rm opt}$ is normalized), 
and that its variance is given by 
\be
\sigma^2_{\rm opt} 
\equiv \langle \hat X_{\rm opt}^2\rangle - \langle \hat X_{\rm opt}\rangle^2
= \frac{1}{\sum_i \lambda_i} 
= \frac{1}{\sum_{i,j}\left(C^{-1}\right)_{ij}}\,.
\ee
{\em Proof}:
The second equality in the above expression for $\sigma_{\rm opt}^2$ 
follows from
\be
\begin{aligned}
\left(\sum\lambda\right)^{-2}
\sum_{i,j} \lambda_i\lambda_j \langle \hat X_i\hat X_j\rangle - X^2
&=\left(\sum\lambda\right)^{-2}
\sum_{i,j} \lambda_i\lambda_j (C_{ij} +\langle \hat X_i\rangle\langle\hat X_j\rangle)-X^2
\\
&=\left(\sum\lambda\right)^{-2}
\sum_{i,j} \sum_{k,l} (C^{-1})_{ik}(C^{-1})_{jl}C_{ij}
\\
&=\left(\sum\lambda\right)^{-2}
\sum_{k,l} (C^{-1})_{kl}
=\left({\sum\lambda}\right)^{-1}\,.
\end{aligned}
\ee
The square of the expected signal-to-noise ratio of 
the optimal estimator is thus
\be
\rho^2_{\rm opt} 
\equiv \frac{\langle \hat X_{\rm opt}\rangle^2}{\sigma_{\rm opt}^2}
= X^2\,\sum_{i,j}\left(C^{-1}\right)_{ij}\,.
\label{e:rho_opt_gen}
\ee
The above construction is a generalization of standard $1/\sigma_i^2$
inverse-noise weighting of $N$ {\em independent} estimators $\hat X_i$, 
which allows for {\em correlations} between the estimators.

\subsection{Optimal common-spectrum estimator}
\label{s:CP}

The optimal linear combination of the common-spectrum 
estimators $\hat A_a^2$, where $a=1,2,\cdots, N$, 
is thus given by
\be
\hat A_{\rm CP}^2 
\equiv \frac{\sum_a \lambda_a \hat A_a^2}
{\sum_b \lambda_b}\,,
\qquad 
\lambda_a\equiv \sum_b \left(C^{-1}\right)_{ab}\,,
\ee
where the subscript `CP' stands for common process,
and $C_{ab}$ are the components of the $N\times N$
covariance matrix 
\be
C_{ab} \equiv \langle \hat A_a^2 \hat A_b^2\rangle 
- \langle \hat A_a^2\rangle
\langle \hat A_b^2\rangle\,.
\ee
To calculate the components $C_{ab}$ of  the 
covariance matrix,
we will need to evaluate expectation values of the form
\be
\langle |\tilde d_{al}|^2|\tilde d_{bl}|^2\rangle\,,
\quad
\langle |\tilde d_{ah}|^2|\tilde d_{bh}|^2\rangle\,,
\quad
\langle |\tilde d_{al}|^2|\tilde d_{bh}|^2\rangle\,,
\quad
\langle |\tilde d_{ah}|^2|\tilde d_{bl}|^2\rangle\,,
\label{e:quartic_CP}
\ee
where $a=b$ for the diagonal elements.
To do so, we will make use of the following identity for
zero-mean Gaussian random variables:
\be
\langle XYZW\rangle
=
\langle X Y\rangle\langle Z W\rangle
+\langle X Z\rangle\langle Y W\rangle
+\langle X W\rangle\langle Y Z\rangle\,.
\label{e:abcd}
\ee
We will also use the following expressions for the various
quadratic expectation values:
\begin{align}
&\langle \tilde d_a(f) \tilde d_b^*(f)\rangle 
= \frac{T}{2}\,P_{ab}(f)\,,
\label{e:dd*}
\\
&\langle \tilde d_a(f) \tilde d_b^*(f')\rangle = 0
\ {\rm if}\ f\ne f'\,,
\label{e:ff'}
\\
&\langle \tilde d_a(f) \tilde d_b(f)\rangle 
=\langle \tilde d_a(f) \tilde d_b^*(-f)\rangle = 0
\ {\rm if}\ f\ne 0\,,
\label{e:dd}
\\
&\langle \tilde d_a^*(f) \tilde d_b^*(f)\rangle 
=\langle \tilde d_a(-f) \tilde d_b^*(f)\rangle = 0
\ {\rm if}\ f\ne 0\,.
\label{e:d*d*}
\end{align}
Note that the last two results hold even if $a=b$.

Using the above results, we find
\be
C_{aa}
= f_l^{26/3}\left(P_{\rm gw}^2(f_l) + 2 \bar\sigma_a^2 P_a(f_l)\right)\,,
\qquad
C_{ab}= 
\left(\bar A_{\rm gw}^2\right)^2\chi_{ab}^2 \,.
\label{e:Cab_CP}
\ee
for the diagonal and off-diagonal elements of the covariance matrix
for the common-spectrum estimators.
These expressions simplify in the weak-signal and intermediate-signal
regimes.
In these limits
\begin{align}
&C_{ab}\simeq 2 f_l^{26/3}(\bar\sigma_a^2)^2\,\delta_{ab}
&({\rm weak\ signal\ regime})\,,
\label{e:Cab_weak}
\\
&C_{ab}\simeq (\bar A_{\rm gw}^2)^2\chi_{ab}^2
&({\rm intermediate\ signal\ regime})\,,
\label{e:Cab_intermediate}
\end{align}
where $\chi_{ab}=1$ in the last expression if $a=b$.
Note that in the weak-signal regime, the covariance matrix is 
diagonal, which means that the individual common-spectrum 
estimators are {\em independent} of one another.
In contrast, in the intermediate-signal regime, 
the covariance matrix has
non-negligible off-diagonal terms corresponding to 
covariances between the estimators, which reduces the 
number of independent terms in the weighted average.

Using \eqref{e:rho_opt_gen}, we can now calculate the square of 
the expected signal-to-noise ratio for the optimal 
combination of common-spectrum estimators
in these two limits.
We just need to invert the above covariance matrices and
sum their components.
To simplify the calculation further we will assume that the 
pulsar white noise power $\bar\sigma_a^2$ is the same 
for all pulsars---i.e., $\bar\sigma_a^2\equiv \bar\sigma^2$.
Then for the $N=45$ pulsars used in NANOGrav's 12.5-year 
analysis, we find
\begin{align}
&\rho^2_{\rm CP} 
\simeq \frac{45}{2} 
P_{\rm gw}^2(f_l)/(\bar\sigma^2)^2 
&({\rm weak\ signal\ regime})\,,
\label{e:rho2_CP_weak}
\\
&\rho^2_{\rm CP} \simeq 20.0
&({\rm intermediate\ signal\ regime})\,.
\label{e:rho2_CP_intermediate}
\end{align}
Note that the factor of $45/2$ in the first expression 
comes from $(C^{-1})_{ab}$ being proportional to the 
identity matrix $\delta_{ab}$ in the weak-signal regime
with $\bar\sigma_a^2=\bar\sigma^2$ for all pulsars.
This is reduced slightly to 20 in the intermediate-signal 
regime, due to the off-diagonal terms in the inverse
covariance matrix.

\subsection{Optimal Hellings-and-Downs estimator}
\label{s:HD}

We now perform a similar calculation for the Hellings and 
Downs cross-correlation estimators $\hat A^2_{ab}$, where
$ab=12, 13, \cdots$.
To evaluate the components $C_{ab,cd}$ of the $N_{\rm pairs}\times N_{\rm pairs}$
covariance matrix between these estimators, we need to evaluate 
expectation values of the form
\be
\langle \tilde d_{al} \tilde d_{bl}^* \tilde d_{cl}\tilde d_{dl}^*\rangle\,,
\quad
\langle \tilde d_{al}^* \tilde d_{bl} \tilde d_{cl}\tilde d_{dl}^*\rangle\,,
\quad
\langle \tilde d_{al} \tilde d_{bl}^* \tilde d_{cl}^*\tilde d_{dl}\rangle\,,
\quad
\langle \tilde d_{al}^* \tilde d_{bl} \tilde d_{cl}^*\tilde d_{dl}\rangle\,,
\label{e:quartic_HD}
\ee
where $a\ne b$, $c\ne d$, but
one or two pulsars might be shared between the
pairs $ab$ and $cd$.
(For example, the diagonal elements of the covariance
matrix have $c=a$ and $d=b$.)
There are thus three cases to consider, and which lead
to the following results:
\begin{enumerate}[(i)]
\item when $ab$ and $cd$ do not share any pulsar in common:
\be
C_{ab,cd} = \frac{1}{2}(\bar A_{\rm gw}^2)^2\,
\left(\frac{\chi_{ac}\chi_{bd} + \chi_{ad}\chi_{bc}}{\chi_{ab}\chi_{cd}}\right)\,,
\label{e:Cabcd_1}
\ee
\item when $ab$ and $cd$ share one pulsar in common, e.g.,
$c=a$:
\be
C_{ab,ad} = \frac{1}{2}(\bar A_{\rm gw}^2)^2\,
\left(1+ \frac{P_a(f_l)}{P_{\rm gw}(f_l)}\frac{\chi_{bd}}{\chi_{ab}\chi_{ad}}\right)\,,
\label{e:Cabcd_2}
\ee
\item when $ab$ and $cd$ share two pulsars in common, e.g.,
$c=a$, $d=b$:
\be
C_{ab,ab} = \frac{1}{2}(\bar A_{\rm gw}^2)^2\,
\left(1 + \frac{P_a(f_l) P_b(f_l)}{P_{\rm gw}^2(f_l)}\frac{1}{\chi_{ab}^2}\right)\,.
\label{e:Cabcd_3}
\ee
\een
These expressions simplify in the weak-signal
and intermediate-signal regimes to
\be
C_{ab,cd} \simeq 0\,,
\qquad
C_{ab,ad} \simeq 0\,,
\qquad
C_{ab,ab} \simeq \frac{1}{2}f_l^{26/3}\,
{\bar\sigma_a^2 \bar\sigma_b^2}/{\chi_{ab}^2}\,
\qquad({\rm weak\ signal\ regime})\,,
\label{e:C_HD_weak}
\ee
and
\be
\left.
\begin{array}{l}
C_{ab,cd} \simeq \frac{1}{2}\left(\bar A_{\rm gw}^2\right)^2\,
\left(\frac{\chi_{ac}\chi_{bd} + \chi_{ad}\chi_{bc}}{\chi_{ab}\chi_{cd}}\right)
\\
C_{ab,ad} \simeq \frac{1}{2}\left(\bar A_{\rm gw}^2\right)^2\,
\left(1+\frac{\chi_{bd}}{\chi_{ab}\chi_{ad}}\right)
\\
C_{ab,ab} \simeq \frac{1}{2}\left(\bar A_{\rm gw}^2\right)^2\,
\left(1+\frac{1}{\chi_{ab}^2}\right)
\end{array}
\quad\right\}
\qquad({\rm intermediate\  signal\ regime})\,.
\label{e:C_HD_intermediate}
\ee

Just as we did for the optimal common-spectrum estimator, we can 
calculate the square of the expected signal-to-noise ratio of
the optimal Hellings-and-Downs estimator in these two limits
by appropriately summing the elements of the inverse 
covariance matrix, $(C^{-1})_{ab,cd}$.
Assuming as before that $\bar\sigma_a^2=\bar\sigma^2$ for 
all pulsars, we find
\begin{align}
&\rho^2_{\rm HD} 
\simeq 2\sum_{ab}\chi_{ab}^2\,{P_{\rm gw}^2(f_l)}/{(\bar\sigma^2)^2}
\approx 66\,{P_{\rm gw}^2(f_l)}/{(\bar\sigma^2)^2}
&({\rm weak\ signal\ regime})\,,
\label{e:rho2_HD_weak}
\\
&\rho^2_{\rm HD} \simeq 4.66
&({\rm intermediate\ signal\ regime})\,,
\label{e:rho2_HD_intermediate}
\end{align}
for the $N=45$ pulsars used in NANOGrav's 12.5-yr analysis.
Note the drastic reduction from 66 to 4.66 in the 
expressions for the squared signal-to-noise ratio in the 
weak-signal and intermediate-signal regimes.
This is due to the presence of non-negligible off-diagonal 
terms in the inverse covariance matrix for the 
intermediate-signal regime, which reduces the number of 
independent terms in the weighted average.

\subsection{Comparison}
\label{s:comparison}

It is now a simple matter to compare the expected signal-to-noise
ratios for the optimal common-spectrum and Hellings-and-Downs
estimators.
Using the results in 
\eqref{e:rho2_CP_weak}, \eqref{e:rho2_CP_intermediate},
\eqref{e:rho2_HD_weak}, \eqref{e:rho2_HD_intermediate},
we find
\begin{align}
&{\rho_{\rm CP}}/{\rho_{\rm HD}}\approx 0.6
&({\rm weak\ signal\ regime})\,,
\label{e:ratio_weak}
\\
&{\rho_{\rm CP}}/{\rho_{\rm HD}}\approx 2.1
&({\rm intermediate\ signal\ regime})\,.
\label{e:ratio_intermediate}
\end{align}
Note that this last result, as well as the values for the
individual signal-to-noise ratios in the intermediate-signal regime 
($\rho_{\rm CP} \approx 4.5$ and $\rho_{\rm HD}\approx 2.2$)
are broadly consistent with NANOGrav's 
recent analysis~\cite{abb+2020} of their 12.5-yr data,
where they found much stronger evidence for a common-spectrum
process than for Hellings-and-Downs cross correlations.

We see that the 
transition between the weak-signal and intermediate-signal 
regimes leads to a reversal of the signal-to-noise preference 
between using the optimal common-spectrum estimator or 
Hellings-and-Downs cross-correlation estimator.
In the weak-signal regime, the noise is dominated by white noise,
and the fact that there are more pairs of pulsars than individual
pulsars means that there is more 
averaging available for the cross-correlation estimator than 
the common-spectrum (i.e., auto-correlation) estimator.
In the intermediate-signal regime, the variance of the GWB itself 
dominates, and because the Earth-term component of the signal 
is correlated across the sky, the number 
of independent terms in the average is considerably less than 
the number of pairs of pulsars, and the benefit of averaging is
reduced by roughly an order of magnitude. 
The common-spectrum estimator also suffers a reduction in the 
number of independent measurements in the intermediate-signal
regime, but this effect is relatively smaller since the 
pulsar-term component of the signal is uncorrelated and the 
number of auto-correlations is smaller. 
As a result, once the variance of the GWB itself dominates, 
as is the case in the NANOGrav 12.5-yr data, the signal-to-noise 
balance begins to favor the auto-correlations.
Figure~\ref{f:SNR_ratio} illustrates the transition between the
weak-signal and intermediate-signal regimes, using the full 
expressions for the covariance matrices \eqref{e:Cab_CP} and
\eqref{e:Cabcd_1}, \eqref{e:Cabcd_2}, \eqref{e:Cabcd_3} to calculate
the expected signal-to-noise ratios $\rho_{\rm CP}$ and $\rho_{\rm HD}$.
This calculation is done for the $N=45$ NANOGrav 12.5-yr pulsars.
\begin{figure}[h!tbp]
\centering
\includegraphics[width=0.5\textwidth]{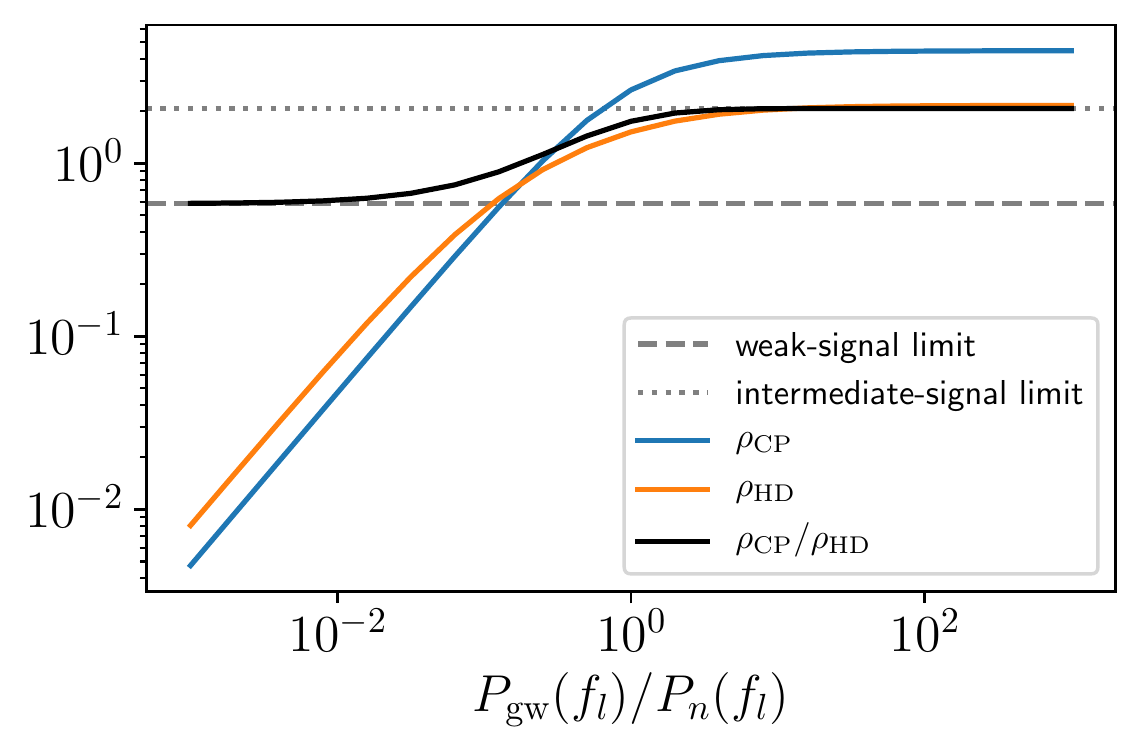}
\caption{Reversal of signal-to-noise preference from $\rho_{\rm HD}$
to $\rho_{\rm CP}$ as one transitions from the weak-signal regime to
the intermediate-signal regime.
The ratios given in \eqref{e:ratio_weak} and \eqref{e:ratio_intermediate}
are represented here by the dashed and dotted horizontal lines.
This figure is for the $N=45$ NANOGrav 12.5-yr pulsars.}
\label{f:SNR_ratio}
\end{figure}
%

\section{Discussion}
\label{s:discussion}

We conclude this paper with a couple of remarks:

(i) The main result---that NANOGrav should see evidence of a GWB 
first in the auto-correlations and then in the 
cross-correlations---is specific to pulsar timing analyses
and is {\em not} expected for ground-based 
detectors like advanced LIGO and Virgo.
This is because the advanced LIGO and Virgo detectors are 
operating in the weak-signal regime, where
$P_{\rm gw}(f)\ll P_n(f)$ for all frequencies.
Hence, for LIGO-Virgo analyses, it is most efficacious to
look for evidence of a GWB in the 
cross-correlations~\cite{Allen-Romano:1999, romano:2016},
where there is 
little~\cite{Schumann_1, Schumann_2, Schumann_3, Schumann_4, mmcs2020}
or no cross-correlated noise to compete against.
In contrast, pulsar timing arrays are most-likely currently 
operating in the intermediate-signal regime~\cite{sejr13},
where the power in the GWB dominates that of the noise at 
low frequencies.
As the total observation time increases, one has access to 
lower frequencies (since $f_{\rm min} \sim 1/T$),
where the power in a GWB with a steep red-noise spectral index 
(e.g., $-13/3$ for SMBBH inspirals) can more easily 
exceed the pulsar noise power
(assuming the noise is white or less red than the GWB), 
and move us into the intermediate-signal regime.
For ground-based detectors, increasing the total observation
time does not allow one to access frequencies below 
$\sim 10~{\rm Hz}$, which are dominated by seismic and 
gravity-gradient noise.

(ii) As mentioned above,
the analyses presented in Secs.~\ref{s:toy_model} and \ref{s:optimal_estimators} 
uses simple frequentist estimators and their expected 
signal-to-noise ratios to
compare the relative likelihood of seeing a GWB in the
auto-correlations versus the cross-correlations.
This approach was chosen for its relative simplicity 
as opposed to its statistical rigor.
A more proper analysis would begin by writing down the likelihood
functions for the various signal+noise models\footnote{A fully 
rigorous pulsar timing analysis also requires that the effects
of the deterministic timing model are taken into consideration when 
searching for gravitational waves \cite{hazboun:2019sc,blandford+1984}.} 
being considered---i.e., (i) noise only, (ii) noise plus a GWB 
that is present only via the auto-correlation terms, and 
(iii) noise plus a GWB that is present in both the 
auto-correlation and cross-correlation terms, 
with the correlation coefficients given by the values of the
Hellings and Downs curve (Figure~\ref{f:HD_curve}).
One would then compare these different signal+noise models via
maximum-likelihood ratios or Bayes factors, to see which model
is preferred by the data.
But since these are the types of analyses that were 
{\em actually performed} by NANOGrav when analyzing 
their 12.5-yr data~\cite{abb+2020}, we did not think that it would be 
particularly illuminating to rederive those statistics here.  
Rather, we thought that a ``back-of-the-envelope" 
illustration that a GWB signal will first appear as a 
common-spectrum process would have value to somebody outside the field.

\section*{Acknowledgements}
\label{s:acknowledgements}

JSH, JDR and XS acknowledge support from the NSF NANOGrav Physics Frontier Center (NSF PHY-1430284).
JDR also acknowledges support from start-up funds from Texas Tech University.
We also thank Michael Lam, Andrea Lommen, Andrew Matas, and Michele Vallisneri for many useful 
comments on an early version of this paper.

\bibliography{refs}

\end{document}